# Graph Theory and Optimization Problems for Very Large Networks

Kamal A. Ahmat
CITY UNIVERSITY OF NEW YORK/Information Technology
New York, USA
Email: kamal.ahmat@live.lagcc.cuny.edu

*Abstract—*

**Graph theory provides a primary tool for analyzing and designing computer communication networks. In the past few decades, Graph theory has been used to study various types of networks, including the Internet, wide Area Networks, Local Area Networks, and networking protocols such as border Gateway Protocol, Open shortest Path Protocol, and Networking Networks. In this paper, we present some key graph theory concepts used to represent different types of networks. Then we describe how networks are modeled to investigate problems related to network protocols. Finally, we present some of the tools used to generate graph for representing practical networks.**

*Keywords*—Complex Networks, Graph Theory, Complex Graphs, Network Models, Network Generators.

## I. INTRODUCTION

Graph theory has been studied by researchers for the past few hundreds of years. Throughout these years, several types of graphs have been studied such as trees, random graphs, directed and undirected graphs, and weighted and unweighted graphs. In the past several decades, researchers have been studying graph theory to understand problems related to communication networks and find appropriate solutions. These problems range from issues related to routing protocols, network management and monitoring, or performance optimization. Interestingly, most of the optimization problems that associate networks with graph theory are either NP-Complete or NP-Hard.

In this paper, we present some key graph theory concepts used to represent different types of networks. Then we describe how networks are modeled to investigate problems related to network protocols emphasizing some optimization problems that have been proven to be intractable. Finally, we present some of the tools used to generate graph for representing practical networks.

## II. BASIC MODEL

When investigating problems related to complex networks, graph theory is often used to represent the network. Due to new network technologies and related emerging issues, graph metrics and notations are becoming more complicated. However, main definitions remain the same. In this section we review standard graph definitions used to represent large network infra structures.

Network topologies vary based on the business logic and functionality. Network devices operating at Data Link Layer communicate through Spanning Tree Protocol (STP) [11]. Thus, the corresponding graph is tree and has no cycles. The scenario changes when multiple spanning trees are assumed. In this case, the graph may contain cycles. At the network layer, the network usually represented as non-tree since network layer protocols such as BGP and OSPF function on all types of graphs. A network $N$ is usually represented by graph $G$ that consists of set of vertices $V$ and links $E$ denoted $G(V, E)$. A link e is denoted by a pair of vertices that are incident to it $e(u, v)$. The link can be undirected or directed based on the problem being addressed and the vertices are generally unordered. In Figure 1, we illustrate three types of graphs.

When considering mobile and sensor networks, the situation changes completely. Since wireless network devices communicate by broadcasting, there is no link between different devices.

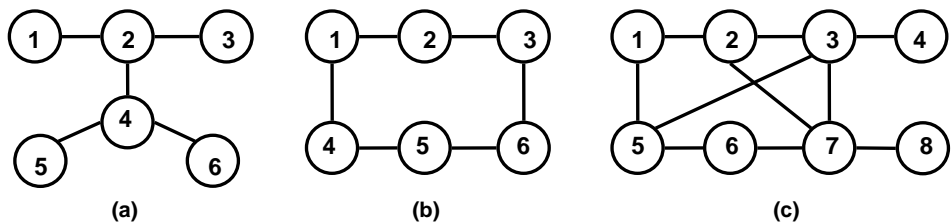

Figure 1: Three types of networks: (a) Tree; (b) Simple network; and (c) Complex network.

III. DIFFERENT NETWORK MODELS AND OPTIMIZATION PROBLEMS

Network models can be viewed in various prospective based on the research topic on interest. In this section, we discuss different views of these network models.

*A. BGP Networks*

The Border Gateway Protocol (BGP) is the routing protocol used to exchange routing information across the Internet [13]. It is used by end users to connect to ISPs and by ISPs connect with each other. BGP is the only protocol that is designed to deal with a network of the Internet's size, and the only protocol that can deal well with having multiple connections to unrelated routing domains.

Internet grows constantly and doesn't have a specific shape. Thus, Internet networks are represented as random graphs. While EBGP is usually used as routing protocol between ASes, IBGP is deployed to exchange routing information between routers within one domain. Optimization problems related to both EBGP and IBGP have been extensively studied as graph theory problems. Links are generally directed and each link in the graph represents a route between ASes or routers within an AS. Unfortunately, many of these problems are proven to be NP-Complete or NP-Hard.

Griffin et al. [8] showed that several BGP optimization problems are NP-Complete. They proved that the detection of oscillations in a SPP instance is NP-Complete through a reduction to 3-SAT [5]. He also showed that Asymmetry and Solvability problems are NP-Complete and Trapped and Unique problems are NP-Hard [8]. Gobjuka [6] further proved that detection of IBGP forwarding loops in NP-Hard but he also proved that safe clustering can be achieved in polynomial time.

### B. Open Shortest Path First

Open Shortest Path First (OSPF) is one of the most commonly used protocols for routing within an AS. Since AS networks consists of hundreds, if not thousands, of subnets, scalability can be an issue. Thus, OSPF supports a two-level hierarchical routing scheme by dividing the network into areas known as OSPF areas. To avoid unnecessary growth of routing tables, addresses that fall within one subnet are aggregated. However, this aggregation has a tradeoff; it can result on loosing information about the shortest path between routers. When studying OSPF, ASes are represented as undirected weighted graphs.

One of problems that have significant practical impact is configuring ASes so that the amount of lost information is minimal when subnets are aggregated. Rastogi et al. [12] studied this problem and proposed an algorithm for optimally configuring an AS when the number of aggregates is fixed. Their algorithm is based on dynamic programming concept and guarantees to minimize the error while computing the shortest path.

### C. Monitoring Schemes

In short, network monitoring is the ability to collect and analyze network traffic. Most intelligent networking devices offer analysis of layer 1 traffic. At this level, the analysis typically focuses on physical network problems such as link status, CRC errors, bipolar violations, and framing errors. Network monitoring has been around as long as there have been networks. Most routers, switches, and intelligent hubs collect some level of network traffic statistics. This information is important to network administrators who are responsible for the operation of the network. Without network monitoring systems, it would be difficult to identify and resolve many network problems.

When studying network monitoring schemes, graph representation of the network can widely vary depending on the specific scheme. When Internet is being monitored, then each vertex in the graph represents an AS. [7] addressed this issue and showed that the optimization problem on minimizing the monitoring cost is NP-Hard. A different scheme has been investigated by Bejerano et al. [1] to monitor administrative domains. Each vertex in such model represents a router and each monitoring tree spans all vertices in the graph. The authors showed that minimizing monitoring overhead is NP-Hard when links are weighted. Breitbart et al. [3] studied the same problem and proved that minimizing the monitoring cost remain NP-Hard when all links are unweighted regardless whether the trees are given or not. In [4], the authors explored the same problem when links are directed and showed that the problem remains NP-Hard. Several other researchers addressed different variations of the problem [9][14].

IV. EXISTING SOFTWARE FOR GENERATING NETWORK GRAPHS

This section describes some graph generating tools that have been used to represent networks. These tools are vendor-independent and based on public standards. We first describe the configuration Model for Power-Law [2] and BRITE, one of the most commonly used Power-Law topology generators. Then we describe Network simulator (ns)[10] explaining its main differences from Power-Law-based models and tools.

*A. Power-Law Model*

A power law can be stated as a polynomial relationship between two quantities. Many types of communication networks exhibit power-law link distributions. That is, in the simplest form, there are few vertices in the network which have a very high degree, and low, degree. And the frequency of vertices that have the average number of degree is very high. Vertices that have very high degree represent central nodes of the network; while vertices with degree one are leaves (end users). Figure 2 illustrates the difference between graphs that exhibit Power-Law model and randomly generated graphs.

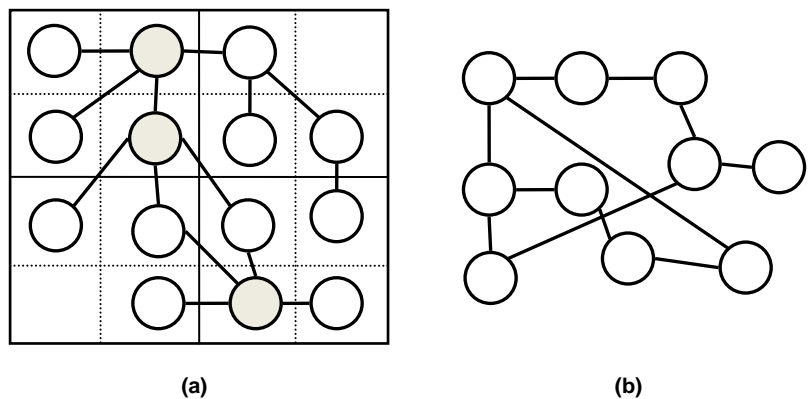

Figure 2: (a) Graph that exhibits Power-Law; (b) Random Graph.

BRITE is one of the most commonly used tools to generate Power-Law-based networks. It divides the plane into squares recursively. Then, it places vertices randomly on the plane. After that, it adds links between vertices based on Power-Law model. BRITE supports multiple generation models including models for flat AS, flat Router and hierarchical topologies. Models can be enhanced by assigning links attributes such as bandwidth and delay.

*B. Network Simuator Model*

Network Simulator (ns) is a discrete event simulator targeted at networking research [10]. Ns provides substantial support for simulating communication protocols such as TCP, routing protocols, and multicast protocols over wired and wireless (both local and satellite) networks. Ns has the capability of simulating live networks by introducing live traffic. This is done by injecting traffic from the simulator into the live network.

The simulator acts like a router allowing real-world traffic to be passed through without being manipulated. The ns packet contains a pointer to the network packet. Network packets may be dropped, delayed, re-ordered or duplicated by the simulator. Opaque mode is useful in evaluating the behavior of real-world implementations when subjected to adverse network conditions that are not protocol specific.

The real-time scheduler implements a soft real-time scheduler which ties event execution within the simulator to real time. Provided sufficient CPU horsepower is available to keep up with arriving packets, the simulator virtual time should closely track real-time. If the simulator becomes too slow to keep up with elapsing real time, a warning is continually produced if the skew exceeds a pre-specified constant ``slop factor'' (currently 10ms).

## V. CONCLUSION

Graph theory has been studied extensively in association with complex communication networks. We described basic concepts of graph theory and their relation to communication networks. Then we presented some optimization problems that are related to routing protocols and network monitoring and showed that many of the optimization problems are NP-Complete or NP-Hard. Finally, we explained some of the common tools used to generate network topologies based on graph theory.